\documentclass[%
 reprint,
superscriptaddress,
 amsmath,amssymb,
 aps,
prb,
floatfix,
]{revtex4-1}

\usepackage{graphicx}
\usepackage{tikz}
\usepackage{xspace}
\usepackage{dcolumn}
\usepackage{bm}
\usepackage[colorlinks=true,allcolors=blue]{hyperref}
\usepackage{array,booktabs,tabularx}
\usepackage[caption=false]{subfig}
\usepackage[USenglish]{babel}
\usepackage{braket}
\usepackage{xcolor}
\usepackage[normalem]{ulem}
\usepackage{notes2bib}
\usepackage{subcaption}
\usepackage{multirow}
\usepackage{array}
\usepackage{alphabeta}
\usepackage[titletoc]{appendix}
\usepackage[utf8]{inputenc}
\usepackage[T1]{fontenc}
\usepackage{float}
\usetikzlibrary{positioning}
\newcolumntype{P}[1]{>{\centering\arraybackslash}p{#1}}
\newcolumntype{M}[1]{>{\centering\arraybackslash}m{#1}}
\usepackage{titlesec}
\titlespacing*{\section}{0pt}{12pt}{6pt}    
\titlespacing*{\subsection}{0pt}{10pt}{5pt} 

\begin{document}
\raggedbottom

\title{A Comprehensive Study on A$_2$PdH$_2$: From Ambient to High Pressure}

\author{Zahra Alizadeh}
\affiliation{Superconductivity Research Laboratory (SRL), Department of Physics, University of Tehran, North Kargar Av., P.O. Box 14395-547, Tehran, Iran}

\author{Yue-Wen Fang}
\affiliation{Centro de F{\'i}sica de Materiales (CFM-MPC), CSIC-UPV/EHU, Manuel de Lardizabal Pasealekua 5, 20018 Donostia/San Sebasti{\'a}n, Spain}

\author{Ion Errea}
\affiliation{Fisika Aplikatua Saila, Gipuzkoako Ingeniaritza Eskola, University of the Basque Country (UPV/EHU), Europa Plaza 1, 20018 Donostia/San Sebasti{\'a}n, Spain }
\affiliation{Centro de F{\'i}sica de Materiales (CFM-MPC), CSIC-UPV/EHU, Manuel de Lardizabal Pasealekua 5, 20018 Donostia/San Sebasti{\'a}n, Spain}
\affiliation{Donostia International Physics Center (DIPC), Manuel de Lardizabal Pasealekua 4, 20018 Donostia/San Sebasti{\'a}n, Spain
}

\author{M.R. Mohammadizadeh}
\email{zadeh@ut.ac.ir}
\affiliation{Superconductivity Research Laboratory (SRL), Department of Physics, University of Tehran, North Kargar Av., P.O. Box 14395-547, Tehran, Iran}

\date{\today}
\begin{abstract}
We present a comprehensive first-principles study of the structural stability and superconducting behavior of Li$_2$PdH$_2$ under high pressure. Using random structure searching and phonon calculations, we identify a pressure-induced phase transition from a tetragonal $I4/mmm$ structure, stable up to $\approx 5$ GPa, to a monoclinic $C2/m$ phase that remains thermodynamically stable up to 50 GPa. Superconductivity is absent in the tetragonal phase, even when anharmonic effects are considered, due to weak electron–phonon coupling and limited hydrogen involvement near the Fermi level. In contrast, the monoclinic phase exhibits a weak but pressure-enhanced superconducting transition, with $T_c$ increasing from 0.6 K at 10 GPa to 4.7 K at 50 GPa, mainly driven by low-frequency Li- and Pd-derived phonon modes. We further explore the isostructural $A_2$PdH$_2$ (A = Na, K, Rb, Cs) series to evaluate the impact of alkali-metal substitution on stability and superconductivity. Na, K, and Rb analogs retain dynamic stability at ambient pressure, with weak superconducting critical temperatures of 3.2 K, 2.1 K, and negligible $T_c$, respectively. Cs$_2$PdH$_2$, however, exhibits phonon instabilities, suggesting a need for external stabilization. These findings highlight the delicate balance between lattice dynamics, electronic structure, and atomic mass in tuning superconductivity in palladium-based hydrides.

\end{abstract}

\maketitle

\section{Introduction} 

Over the past decade, advances in high-pressure synthesis and characterization techniques have enabled the discovery of a diverse range of metal hydrides with remarkable physical properties \cite{bronger1998new, bronger1995high, rusman2016review, modi2021room, duan2017structure}.
These materials, encompassing both binary and ternary systems, exhibit remarkable versatility under extreme conditions, enabling applications across hydrogen storage, catalysis, and quantum materials research.
Among their diverse functionalities, the potential for superconductivity has drawn particular interest. Several binary hydrides have been experimentally shown to exhibit superconducting behavior under pressure, with critical temperatures ($T_c$) exceeding 200 K. Notable examples include H$_3$S ($T_c$ $\approx$ 203 K at 150 GPa) \cite{drozdov2015conventional}, LaH$_{10}$ ($T_c$ up to 260 K at $\approx$ 170 GPa) \cite{Drozdov2019-La-H-Nature, somayazulu2019evidence}, CaH$_6$ ($T_c$ $\approx$ 210 K at $\approx$ 160–170 GPa) \cite{ma2022high}, YH$_9$ ($T_c$ $\approx$ 243–262 K at $\approx$ 180–200 GPa) \cite{du2023superconducting}, and AcH$_x$ \cite{semenok2018actinium}. These compounds typically feature dense hydrogen networks and strong electron–phonon coupling, which contribute to their unique quantum behaviors. This has positioned high-pressure hydrides as one of the most promising platforms for exploring high-temperature superconductivity.

While both binary and ternary metal hydrides have demonstrated remarkable versatility under high pressure, their superconducting behaviors often prove to be highly sensitive to subtle changes in stoichiometry and crystal structure. For instance, compounds like LiPdH are metallic yet non-superconducting \cite{singh1990possiblity, alizadeh2025superconductivity}. In particular, LiPdH has been found to remain both thermodynamically and mechanically stable under pressures up to 100 GPa, yet without exhibiting superconductivity. In contrast, a modification into Li$_2$PdH$_6$—featuring a different hydrogen-to-metal ratio and crystal framework—leads to theoretical predictions of superconductivity with critical temperatures as high as 165 K at 90 GPa, with superconductivity persisting down to 10 GPa ($T_c$ $\approx$ 106 K) \cite{xia2024chemically}.
This highlights how atomic rearrangements can strongly influence bonding character, lattice dynamics, and ultimately, the electron–phonon coupling strength.

In this context, ternary hydrides offer an expanded compositional space that allows for finer control over subtle structural and electronic variations, especially under high pressure. Among them, transition metal-based hydrides exhibit a rich interplay of ionic, covalent, and metallic bonding, which can lead to unusual quantum phenomena, phase transitions, and strongly coupled lattice dynamics. Their complex bonding environments make them ideal candidates for probing pressure-induced emergent behavior and for designing materials with highly tunable properties. This structural and electronic versatility originates from their flexible chemical composition. A new class of ternary hydrides can be created by combining different metal hydrides under suitable conditions.
 In the synthesis of compounds with the general formula $A_xT_y$H$_z$:
\begin{equation}
   xA {\rm H} + yT + ((z - x)/2) {\rm H}_2 \rightarrow A_xT_y{\rm H}_z, 
\end{equation}
one metal ($A$) is typically derived from the alkali or alkaline earth metal hydrides, while the other metal ($T$) comes from transition metals. The resulting compounds exhibit diverse electronic characteristics—ranging from insulating and semiconducting to metallic and superconducting properties—depending on their composition and structural arrangement \cite{ronnebro2003gigapascal, puhakainen2010synthesis, bronger1995high, rusman2016review, modi2021room, duan2017structure}. Among transition metals, palladium (Pd) is particularly notable for its exceptional properties. Under ambient conditions, Pd can absorb a remarkable amount of hydrogen, making it an ideal candidate for hydrogen-related technologies \cite{alizadeh2025superconductivity, zhao2024superconducting, saha2023mapping, liu2023nanostructured, klopvcivc2023review, gillions2023recent, xu2020nanoporous, yamauchi2008nanosize}. Beyond the well-studied binary PdH$_x$ phases, a growing number of ternary $A$–Pd–H systems have been proposed or synthesized, featuring various alkali metals as the $A$-site element \cite{meninno2023abinitio,sakamoto1995hydrogen, bronger1995hochdrucksynthese, bronger1998new, vocaturo2022prediction, ronnebro2003gigapascal, frost2022high, bronger1995high, wang2024advances}.  These ternary compounds open new avenues for tuning physical properties, including superconductivity, through chemical composition and structural configuration. Several stoichiometries involving different alkali metals—such as Cs$_2$PdH$_4$ and Cs$_3$PdH$_5$ \cite{bronger1992darstellung}, Rb$_2$PdH$_4$ \cite{bronger1992darstellung}, K$_2$PdH$_4$ \cite{kadir1991metallic}, and Na$_2$PdH$_2$, NaPd$_3$H$_2$ and Na$_2$PdH$_4$ \cite{bronger1998new, kadir1993structure, kadir1991metallic, ronnebro2003gigapascal} and Li$_2$PdH$_2$ \cite{kadir1989li2pdh2, yao2017high}, LiPdH \cite{singh1990possiblity, noreus1990absence, Liu2017, alizadeh2025superconductivity}—have been either experimentally synthesized or theoretically proposed in the literature. While structural data exist for several $A$–Pd–H compounds, their electronic and superconducting characteristics are still poorly understood, due to the absence of comprehensive experimental or theoretical analyses. Most of these compounds exhibit insulating behavior; however, a few, such as Li$_2$PdH$_2$, LiPdH and Na$_2$PdH$_2$, have been reported to show metallic characteristics, though no experimental evidence of superconductivity has yet been observed in these phases. Li$_2$PdH$_2$—reported following the discovery of Na$_2$PdH$_2$ \cite{noreus1988na2pdh2}—has garnered significant interest due to its metallic nature and unique properties \cite{kadir1989li2pdh2}. Its structure closely resembles that of Na$_2$PdH$_2$, adopting a tetragonal configuration with linear [Li–H–Pd–H–Li] complexes, and exhibiting only a slight lattice contraction due to the smaller radius of the lithium atom \cite{kadir1989li2pdh2}. Given the sensitivity of metal hydrides to external pressure and the possibility of structural and electronic transitions under compression, investigating their high-pressure behavior becomes essential for understanding their functional potential. In this context, prior work on the deuterated analog Li$_2$PdD$_2$ revealed possible pressure-induced phase transitions and notable changes in electronic structure up to 50 GPa \cite{yao2017high}. Motivated by these findings—and the lack of detailed electronic and superconducting characterization—we select Li$_2$PdH$_2$ as a benchmark compound for comprehensive theoretical analysis.

In this work, we systematically investigate the structural stability and superconducting behavior of Li$_2$PdH$_2$ over a wide pressure range (0–50 GPa). In the first part, we employ random crystal structure searching to determine the thermodynamically stable phases, identifying a tetragonal $I4/mmm$ phase at low pressure and a monoclinic $C2/m$ phase stabilized above $\approx 5$ GPa. In the second part, we focus on the superconducting behavior of both phases, analyzing the electron–phonon coupling with and without the consideration of anharmonicity. In the final part of the study, we systematically extend the analysis to isostructural compounds $A$$_2$PdH$_2$ ($A$ = Na, K, Rb, Cs) to evaluate how alkali substitution influences structural stability, electronic structure, and superconducting characteristics. This uniform framework allows us to evaluate the influence of alkali substitution on structural stability, electronic structure, and electron–phonon coupling, ultimately guiding the search for new Pd-based hydrides with enhanced superconducting properties under moderate pressures.


\section{Computational Methods}

The crystal structures of Li$_2$PdH$_2$ were explored across a pressure range from ambient up to 50 GPa using fixed-composition structural searches, which were conducted using the CRYSPY code \cite{Yamashita2021}. For these calculations, 20 generations were used, each containing 50 candidate structures. The first generation was initialized with randomly generated structures. While CRYSPY handles the structural evolution, {\sc Quantum ESPRESSO} \cite{Giannozzi2009, Giannozzi2017}, with a plane-wave cutoff energy of 100 Ry, was utilized for optimizing the geometry and calculating enthalpies. The Brillouin zone sampling in the crystal structure prediction was determined automatically based on each structure’s symmetry and lattice parameters, ensuring an adaptive k-point density. The electron-ion interaction was described using ultrasoft pseudopotentials, considering ${5s}^1$ and ${4d}^9$ electrons included in the valence of Pd. All total energy calculations and structural relaxations were carried out within the framework of density functional theory (DFT), using the generalized gradient approximation (GGA) in the form of the Perdew–Burke–Ernzerhof (PBE) parameterization for the exchange-correlation functional \cite{perdew1996generalized}.

We also employed the {\sc Quantum ESPRESSO} package to compute the electronic and phononic properties of the relaxed crystal structures. 
We employed the same pseudopotentials and exchange-correlation functionals in the electronic and phononic calculations as in the crystal structure predictions to ensure consistency. Lattice dynamics and electron-phonon coupling were investigated using density functional perturbation theory (DFPT) \cite{giannozzi1991ab}. Electronic properties, such as the electron localization function (ELF) \cite{Becke1990} and Bader charge \cite{bader1990}, were computed using the postprocessing tools provided by {\sc Quantum ESPRESSO} and the crystal orbital 
Hamiltonian population (COHP) was calculated using the LOBSTER  code \cite{maintz2016lobster}. The plane-wave cutoff of 140 Ry and 1000 Ry were set for the wave functions and the charge density, respectively. Brillouin zone integration in the self-consistent calculations utilized a first-order Methfessel-Paxton \cite{Methfessel1989} smearing of 0.02 Ry broadening along with the 18 $\times$ 18 $\times$ 18 ( 24 $\times$ 18 $\times$ 12) k-point grid and dynamical matrices were calculated on a 6 $\times$ 6 $\times$ 6 (8 $\times$ 6 $\times$ 4) q-point mesh for the $I4/mmm$ ($C2/m$) phase.

Next,  stochastic self-consistent harmonic approximation  (SSCHA) \cite{Monacelli2021, Errea2013, Errea2014} computations were performed on a 2 × 2 × 2 supercells for lattice anharmonic corrections on the phonon spectra. To bridge the gap between harmonic and anharmonic dynamical matrices, interpolation to a finer 6 × 6 × 6 grid was performed. Furthermore, the electron–phonon interaction was investigated through both harmonic and quantum anharmonic calculations, employing a 36 $\times$ 36 $\times$ 36 (40 $\times$ 36 $\times$ 32) k-point grid for the $I4/mmm$ ($C2/m$) structure. The critical temperature for superconductivity was determined by the McMillan Allen-Dynes modified equation \cite{allen1975transition}, with ${\mu}^*= 0.1$.

\section{Results}

\subsection{Stability of Li$_2$PdH$_2$}

\begin{figure}
\begin{center}
\includegraphics[width=8.5cm]{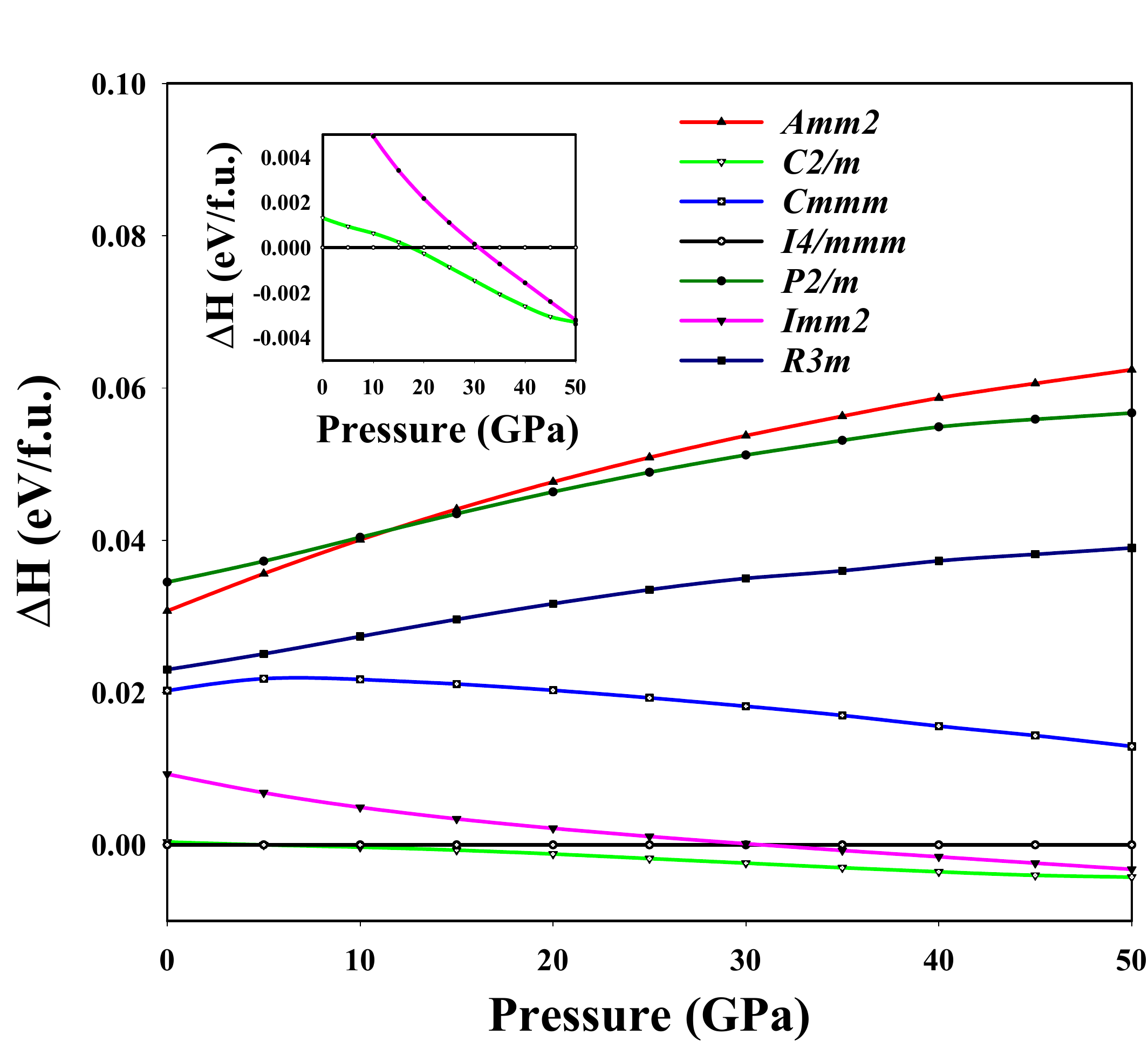}
\caption{Theoretical enthalpy versus pressure for different structures of Li$_2$PdH$_2$. The inset includes the zero-point energy contribution for the $I4/mmm$ and $C2/m$ structures. The enthalpy of the $I4/mmm$ structure is the reference enthalpy.}
\label{Fig1}
 \end{center}
\end{figure}

To explore the pressure-induced structural evolution of Li$_2$PdH$_2$, we performed a crystal structure prediction based on random structure search methods. This approach allows for identification of energetically favorable phases without assuming any predefined transition pathway. The goal is to determine the ground-state structures of Li$_2$PdH$_2$ as a function of pressure and to resolve discrepancies between previous theoretical predictions and experimental observations \cite{yao2017high}.
As illustrated in Fig. \ref{Fig1}, at ambient pressure the tetragonal $I4/mmm$ structure emerges as the most stable phase, consistent with previous experimental findings \cite{kadir1989li2pdh2, yao2017high}. A monoclinic $C2/m$ structure was also identified with an extremely small energy difference of ~0.4 meV/f.u., implying that, at ambient pressure, the energy difference between the two structures is negligible, making them effectively equivalent from a thermodynamic standpoint. Upon increasing pressure, $C2/m$ gradually becomes more favorable, with a marginal enthalpy advantage appearing around 5 GPa. To assess the influence of quantum nuclear effects on phase stability, we incorporated zero-point energy (ZPE) corrections into our enthalpy calculations. The results reveal that, although the enthalpy difference between the $I4/mmm$ and $C2/m$ phases is initially minimal (0.4 meV/f.u.), inclusion of ZPE increases this value to 1.0 meV/f.u. at ambient pressure. This subtle enhancement suggests that even weak quantum fluctuations, primarily associated with hydrogen, can affect the thermodynamic landscape. More notably, the pressure at which the phase transition occurs is also shifted significantly—from approximately 5 GPa to around 17 GPa—when ZPE is taken into account. For clarity, the inset of Fig.~\ref{Fig1} presents the ZPE-corrected enthalpy difference between the two lowest-energy structures (I4/mmm and C2/m), illustrating how the inclusion of quantum effects slightly alters their relative stability, particularly near the transition point.

To benchmark our predictions, we compared the results with the theoretical study by Yao et al. \cite{yao2017high}, which focused on Li$_2$PdD$_2$, the deuterated analogue of our system. Although the systems differ by isotope, the comparison is valid in the context of ground-state energetics and structural stability, which are not significantly affected by isotopic substitution. In Yao’s study, a series of metadynamics simulations were conducted to explore pressure-induced phase transitions. Starting from the tetragonal $I4/mmm$ phase, a low-enthalpy orthorhombic $Pnma$ structure was obtained around 15 GPa,  which did not fully match the experimental XRD patterns. To further refine the structural model, simulations were re-initiated from the $Pnma$ structure, eventually yielding a monoclinic $C2/m$ phase that matched the experimental data closely and was identified as the thermodynamic ground state above $\approx 4.6$ GPa \cite{yao2017high}. In contrast, our structure search reveals a direct transition from $I4/mmm$ to $C2/m$, with no appearance of low enthalpy of the $Pnma$ phase. The Pnma structure, which emerged from our structure search, exhibits a significantly higher enthalpy than 
$I4/mmm$ by more than 0.25 eV/f.u. at 0 GPa with the difference increasing further under pressure, indicating its thermodynamic unfavorability throughout the investigated range.

The $C2/m$ structure becomes favorable already at 5 GPa, aligning with the experimentally observed transition pressure \cite{yao2017high}. This suggests a simpler phase transition pathway in Li$_2$PdH$_2$ compared to the previous study, while still arriving at the same high-pressure phase.
In addition to the $I4/mmm$ and $C2/m$ phases, several structures predicted in our study overlap with those reported previously for the deuterated compound. Notably, the $Imm2$ phase appears in both works as a competitive structure. In our calculations, $Imm2$ becomes energetically comparable to $I4/mmm$ at around 30 GPa, whereas in the previous study, it was found to be favorable near 12 GPa.
The $P2/m$ phase, which was found to be more stable than the $I4/mmm$ structure above 15 GPa in the earlier study, was also identified in our calculations. However, in our case, $P2/m$ remains consistently higher in energy throughout the entire pressure range and does not play a significant role in the structural evolution of Li$_2$PdH$_2$.
Our structure search also revealed several new structural motifs, such as $Cmmm$, $R3m$, and $Amm2$, but they are not energetically favorable in the studied pressure range. While these phases do not compete directly with the lowest-energy configurations, their identification broadens the configurational space of Li$_2$PdH$_2$ and may be of interest for future studies, particularly the $Cmmm$, exploring metastable or temperature-driven phases under higher pressures. 

\subsection{$I4/mmm$ and $C2/m$ phases}

To gain a deeper understanding of the stability and key properties of the predicted phases, we focus on the two most competitive structures identified in our calculations: the $C2/m$ and $I4/mmm$ phases. Given their close proximity in energy to the ground-state configuration, a detailed analysis of their structural, electronic, and vibrational properties is performed. In particular, we assess their dynamical stability under pressure and investigate the possibility of pressure-induced superconductivity, with the aim of clarifying their potential relevance to the phase diagram of Li$_2$PdH$_2$.

At ambient pressure, Li$_2$PdH$_2$ crystallizes in a body-centered tetragonal unit cell ($I4/mmm$), as confirmed by experimental studies \cite{yao2017high, kadir1989li2pdh2}. Our computational study builds on these experimental findings, extending the investigation to higher pressures and examining the stability of Li$_2$PdH$_2$ at various pressures, as shown in Fig. \ref{Fig1}. Our results indicate that upon increasing pressure, the tetragonal phase remains stable up to 5 GPa. The relaxed lattice parameters at ambient pressure align well with experimental values, confirming the accuracy of our computational approach. In the $I4/mmm$ structure, the unit cell parameters are defined as $a = b = 3.08$ \r{A} and $c = 10.24$ \r{A}, which reflects a high degree of symmetry characteristic of the body-centered tetragonal structure. 

\begin{figure*} [htbp]
\begin{center} 
\includegraphics[width=0.9\textwidth]{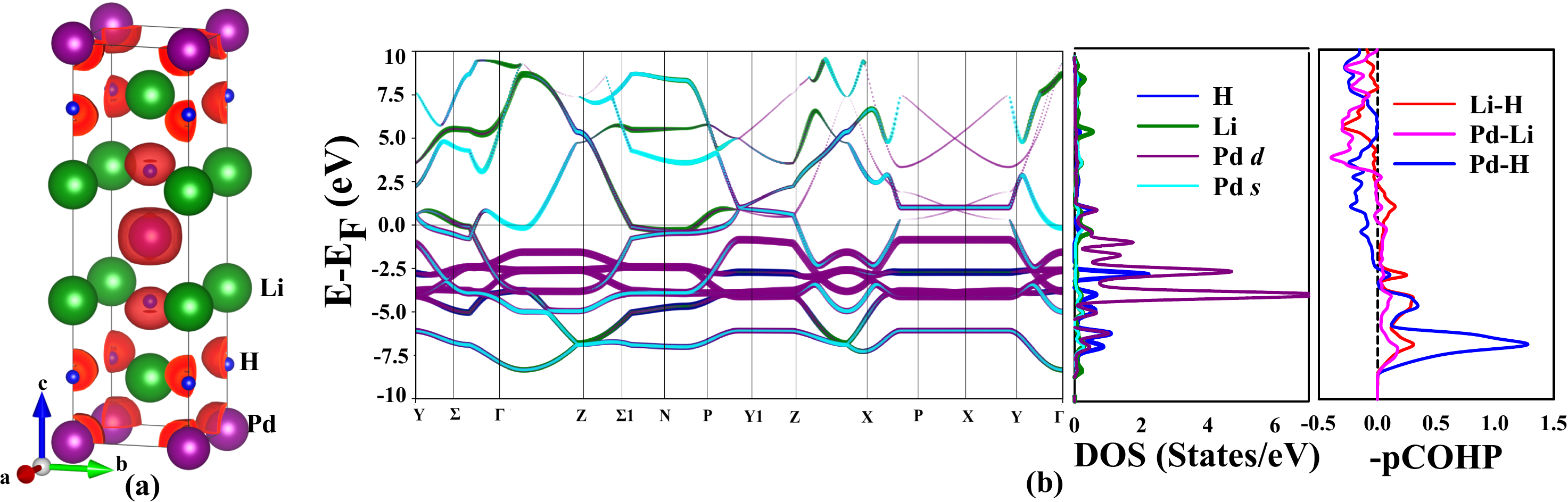}
\vspace{1em}
\includegraphics[width=0.9\textwidth]{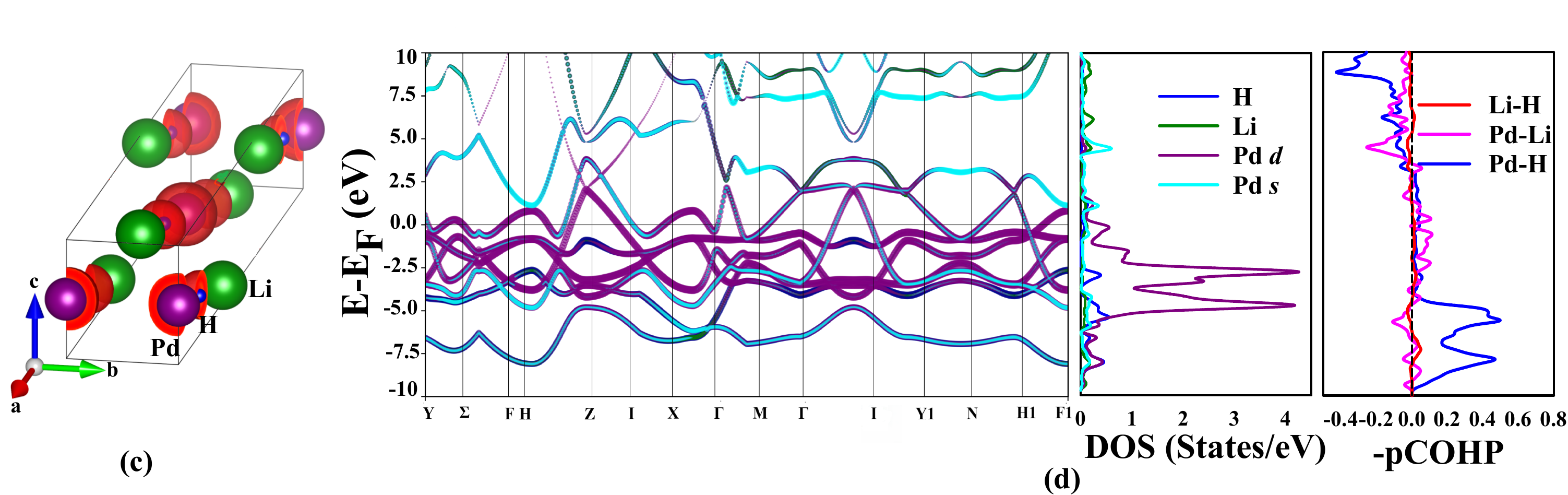}
\caption {(a) The 3D structure, with the corresponding Electron Localization Function plotted on the structures at an isovalue of 0.8 for $I4/mmm$ Li$_2$PdH$_2$ (Higher values of the ELF are depicted in red) (b) The electronic band structure (left panel), density of states (DOS) (middle panel), and -pCOHP for pairs of H-Pd, Li-Pd and Li-H of $I4/mmm$. Accordingly, (c) and (d) correspond to the ELF, electronic structure, and -pCOHP of $C2/m$. The Fermi level is set to zero in panels (b) and (d).}
\label{Fig2} 
 \end{center} 
\end{figure*}

The conventional unit cell of the $I4/mmm$ structure is illustrated in Fig. \ref {Fig2} (a). In this structure, Pd atoms occupy the Wyckoff \textit{a} position (2 sites with $4/mmm$ symmetry), while Li and H atoms occupy the Wyckoff \textit{e} position (4 sites with $4mm$ symmetry). The $I4/mmm$ structure consists of a network of connected pyramids and octahedra. Each Li atom is bonded to four Pd atoms at a distance of 2.56 {\AA} and five H atoms, with one Li–H bond measuring 1.9 {\AA} and the remaining four at 2.19 {\AA}. Pd atoms are coordinated by eight Li and two H atoms, with Pd–H bonds measuring 1.79 {\AA}. Each H atom connects to five Li atoms and one Pd atom, forming octahedral units. This structure consists of $\text{Li}^+$ ions and $\text{[PdH}_2\text{]}^{2-}$ complexes arranged along the $c$-axis.
Structural stability arises from electrostatic interactions between $\text{Li}^+$ cations and terminal H atoms in the $\text{[PdH}_2\text{]}^{2-}$ units, forming linear chains. These chains are further held together through weak Pd–Pd interactions, resulting in an average intermetallic distance of ~3.3 {\AA}. To further investigate the bonding nature between Pd and H atoms in Li$_2$PdH$_2$, we performed crystal orbital Hamilton population (COHP) analysis. As shown in the right panel of Fig.~\ref{Fig2} b, the Pd–H interactions exhibit pronounced bonding features in the energy range between 7.5 and 5 eV below the Fermi level, indicating the presence of covalent bonding. These results confirm that, similar to LiPdH, hydrogen atoms in Li$_2$PdH$_2$ also contribute significantly to the electronic structure by forming directional Pd-H bonds.

Upon transitioning to the monoclinic $C2/m$ phase, the symmetry is significantly reduced, leading to noticeable lattice distortions. The unit cell parameters change to $a = 6.01$ \r{A}, $b = 3.84$ {\AA}, and $c = 2.61$ {\AA}. These changes are accompanied by off-diagonal components in the cell matrix, indicating a monoclinic distortion. In the $C2/m$ structure in Fig. \ref{Fig2} (c), Pd atoms occupy the Wyckoff $2i$ position (2 sites with $2/m$ symmetry), while Li and H atoms occupy the Wyckoff $4g$ position (4 sites with $mm$ symmetry). In this structure, the $\text{[PdH}_2\text{]}^{2-}$ groups form extensive chains with closely associated Pd atoms. In this configuration, Pd–Pd distances shorten to approximately 2.8 \r{A} and strengthening the Pd–Pd interactions. A similar bonding behavior is observed in the high-pressure $C2/m$ phase, as shown in the corresponding COHP plot (Fig.\ref{Fig2} d). The Pd-H bonding states remain prominent, although the bonding energy range slightly extends down to -8~eV, indicating a comparable covalent interaction pattern between Pd and H atoms.

Bader charge analysis shows that each Li atom loses about 0.84$e^-$, while Pd and H atoms gain approximately 0.63$e^-$ and 0.52$e^-$, respectively, in the $I4/mmm$. In the $C2/m$ structure, Li atoms lose a similar amount (0.82$e^-$), while Pd and H atoms gain around 0.56$e^-$ and 0.54$e^-$, respectively. The electron localization functions (ELF) in Fig. \ref{Fig2} (a) and (c) further illustrate this charge transfer and localization. High electron localization forms around Pd and H (red circular regions), indicating their electron-gaining nature due to electronegativity. The Li atoms, shown in blue, exhibit low electron localization, consistent with their role as electron donors.

The electronic band structure and density of states (DOS) for both configurations are shown in Fig. \ref{Fig2} (b) and (d). Unlike typical metallic systems, the Fermi level does not fall within a region of high electronic density; instead, it is situated near the bottom edge of a pseudogap. This positioning reflects an intermediate electronic character, in which conductivity is strongly suppressed. The Pd–H bonding within the PdH$_2$ units is evident in the energy range near 6eV (-7.5eV) below the fermi energy for $I4/mmm$ ($C2/m$).  The DOS near the Fermi level is dominated almost entirely by the Pd $d$ orbitals, indicating that conduction-related states originate primarily from the $\text{PdH}_2$ units. The contribution of Li atoms, as electron donors, to the DOS in the valence region is minimal. The contributions of H, Li, and Pd at the Fermi level for $I4/mmm$ ($C2/m$) are approximately 12$\%$ (19$\%$), 34$\%$ (26$\%$), and 54$\%$ (55$\%$), respectively.

\begin{figure*}
\begin{center} 
\includegraphics[width=2.0
\columnwidth]{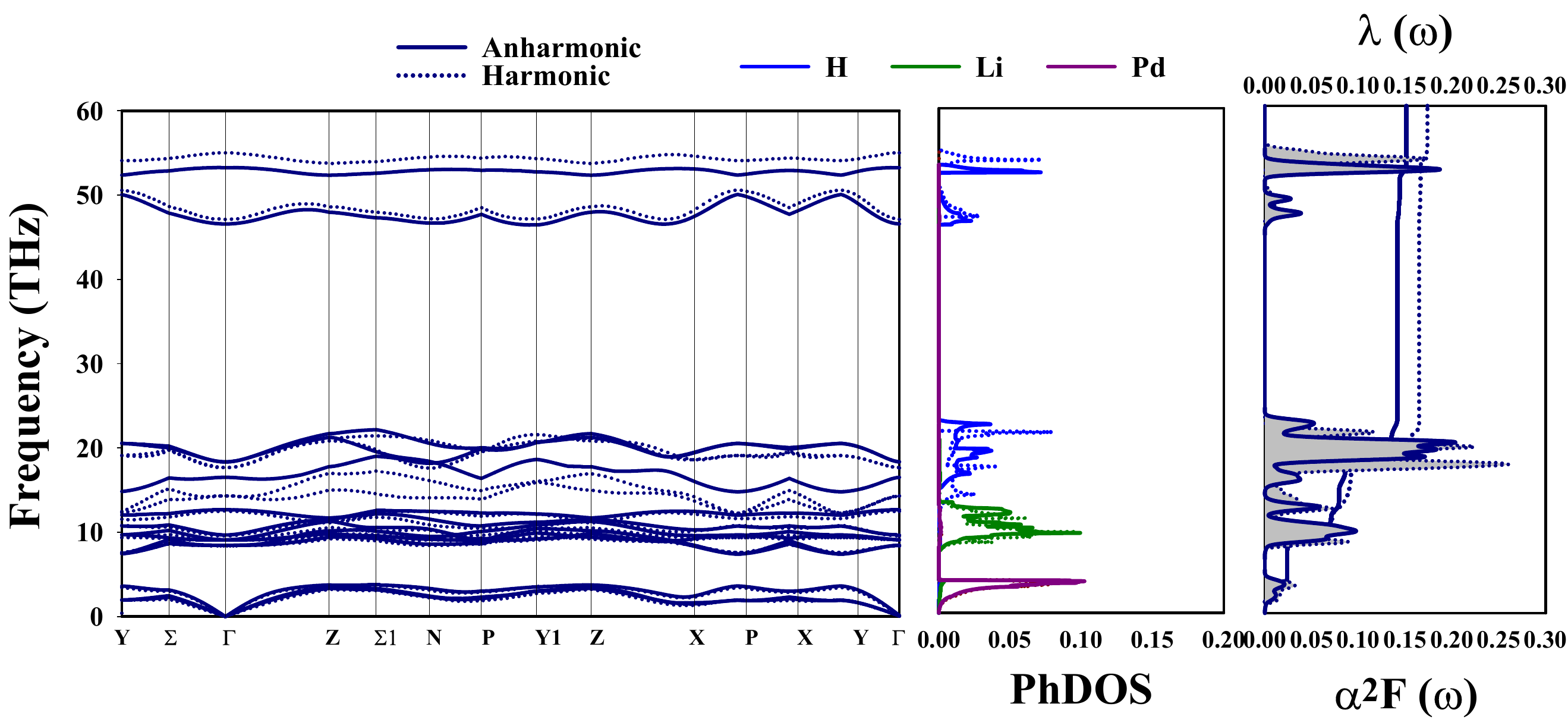} 
\caption{Phonon spectrum (left) and phonon density of states (PDOS) with the Eliashberg function $\alpha^{2}F(\omega)$ and integrated electron-phonon coupling constants $\lambda(\omega)$ in harmonic and anharmonic calculations for $I4/mmm$ Li$_2$PdH$_2$, obtained using SSCHA relaxation.}
\label{Fig3} 
\end{center} 
\end{figure*}

\begin{figure*}
\begin{center}
\includegraphics[width=2.0\columnwidth,draft=false]{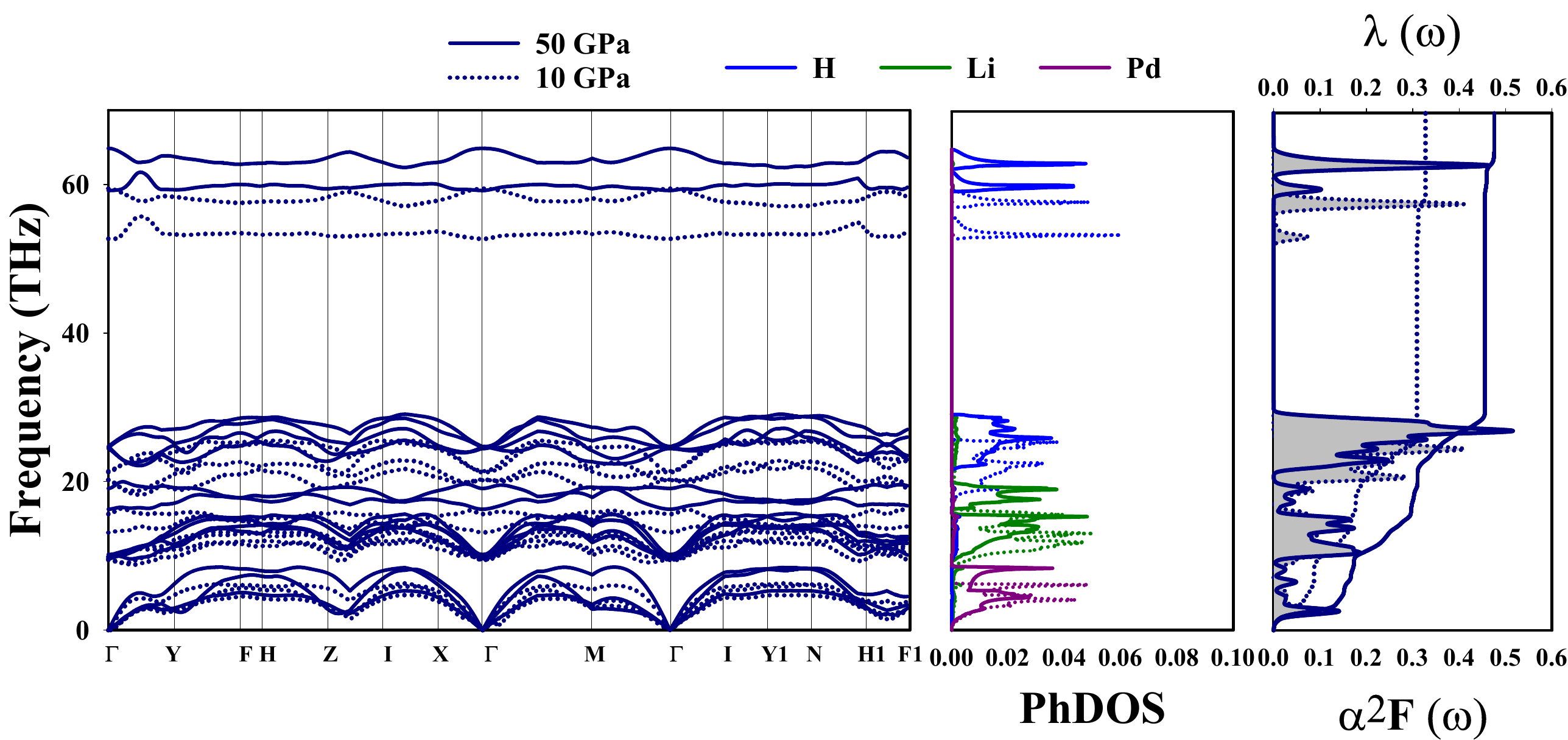}
\caption{The phonon spectrum (left), the phonon density of states (PDOS) (middle), and Eliashberg function $\alpha^{2}F(\omega)$ and integrated electron-phonon coupling constants $\lambda(\omega)$ (right) at 10 and 50 GPa in harmonic approximation for $C2/m$ Li$_2$PdH$_2$.}
\label{Fig4}
\end{center}
\end{figure*}

We conducted an in-depth analysis of the phonon dispersion, the phonon density of states (PhDOS), Eliashberg function ${α}^2F(ω)$, and electron-phonon coupling (EPC) constant λ(ω) for both the tetragonal $I4/mmm$ and monoclinic $C2/m$ phases, as depicted in Fig. \ref{Fig3} and Fig. \ref{Fig4}. For the $I4/mmm$ structure, high-frequency optical phonon modes, occurring above approximately 15 THz, are primarily associated with hydrogen vibrations, while the low-energy acoustic modes (below 5 THz) and the optical modes in the range of 5–15 THz are mainly attributed to Pd and Li atoms, respectively. This atomic distribution across the phonon spectrum is further reflected in the Eliashberg function and the λ(ω), revealing a significant contribution from H-dominated phonon modes. ${α}^2F(ω)$ analysis shows that the majority of the total λ originates from H and Li atoms, contributing approximately $\approx50\%$ and $\approx30\%$, respectively. In contrast, palladium atoms contribute only about $\approx20\%$, with their influence limited to the low-frequency range, where a narrow peak is observed in $\alpha^{2}F(\omega)$.

To account for potential anharmonic effects—which can substantially impact lattice dynamics and superconductivity in hydrides, we performed anharmonic phonon calculations using the SSCHA. As expected, the anharmonic effects are most pronounced in hydrogen vibrations, while their influence on Li-related modes is limited, and negligible for Pd atoms due to the more harmonic nature of their restoring forces. The distinct anharmonic response of the two sets of hydrogen optical modes can be attributed to their different interatomic force characteristics and phonon eigenvector patterns. Lower-frequency H modes (15–22 THz), which strongly contribute to electron–phonon coupling, typically involve larger collective displacements and exhibit relatively shallow local potential wells, making them more prone to stiffening under anharmonic corrections. In contrast, higher-frequency H modes (45–55 THz) often involve more localized, high-energy bond-stretching vibrations, which are more sensitive to anharmonicity and tend to soften.

We also examined the phonon and superconducting properties of the monoclinic phase at 10 GPa, slightly above the transition pressure to ensure full stabilization of the $C2/m$ phase, and 50 GPa. Compared to the $I4/mmm$ phase, the monoclinic structure exhibits a slightly broader phonon frequency range at 50 GPa. Notably, the vibrational energy ranges of both Pd and Li atoms increase in the monoclinic phase, leading to enhanced contributions to the electron–phonon coupling. Specifically, Pd and Li contribute $\approx27\%$ and $\approx53\%$ to the total $\lambda$, respectively, while hydrogen accounts for only about $\approx20\%$. These differences highlight the distinct lattice dynamics and coupling mechanisms between the two structural phases. By solving the Allen-Dynes-modified McMillan equation, we estimate the superconducting transition temperature $T_c$ in the $I4/mmm$ and $C2/m$. The $I4/mmm$ phase shows no superconducting transition under both harmonic and anharmonic conditions, with $T_c$ calculated to be effectively zero. In contrast, the $C2/m$ phase shows an enhancement in superconducting behavior, with $T_c$ increasing from 0.6 K at 10 GPa to 4.7 K at 50 GPa (for $\mu^* = 0.1$). This enhancement can be attributed to a larger $\lambda$ and a higher density of states at the Fermi level in the $C2/m$. Although the hydrogen contribution to $\lambda$ is lower in the $C2/m$ phase ($\approx20\%$) compared to the $I4/mmm$ ($\approx50\%$), the stronger coupling from Pd and Li atoms, compensates for this and enables superconductivity to emerge.

\subsection{$A_2$PdH$_2$: $A$ = Na, K, Rb and Cs}

Following the identification of metallic behavior in Li$_2$PdH$_2$ , we extended our analysis to the isostructural series $A_2$PdH$_2$ (A = Na, K, Rb, Cs), aiming to assess whether similar properties persist across heavier alkali substitutions. Although alternative stoichiometries involving these elements have been reported \cite{bronger1992darstellung, bronger1998new, kadir1993structure, kadir1991metallic, ronnebro2003gigapascal, kadir1989li2pdh2, yao2017high, singh1990possiblity, noreus1990absence, Liu2017, alizadeh2025superconductivity}, they are typically associated with insulating states and were thus not pursued here. Instead, the $A_2$PdH$_2$ composition was selected as a candidate for retaining metallicity while enabling a systematic evaluation of structural, electronic, mechanical stability and superconductivity across the alkali-metal series.

Structural relaxation of the $A_2$PdH$_2$ compounds with heavier alkali metals ($A$ = Na, K, Rb, Cs) reveals a systematic expansion of the unit cell, consistent with the increasing ionic radii. The lattice parameters, extracted from relaxed structures, increase from 5.08 Å (Li) to 6.40 Å (Cs), reflecting a nearly linear trend with $A$-site substitution. Importantly, the crystal symmetry remains preserved, and no significant distortion of the [PdH$_2$] subunits is observed. This suggests that in the $A_2$PdH$_2$ structural framework, the larger ions fit within the structure while reducing internal stresses and maintaining structural stability. Consequently, bond lengths adjust to each substitution’s atomic radius: for Na, K, Rb, and Cs substitutions, the Pd–H bond lengths decrease to approximately 1.98 \text{\AA}. The H-$A$ and Pd-$A$ bond lengths adjust to approximately 1.92, 2.59 \text{\AA} (Na), 2.68, 3.23 \text{\AA} (K), 2.83, 3.37 \text{\AA} (Rb), 2.95, 3.54 \text{\AA} (Cs). Minor shifts in Pd–Pd distances occur as the electronic structure adjusts to the redistributed charge, while the structural integrity of the [PdH$_2$] units remains intact.

Charge distribution analysis indicates that each Na, K, Rb, and Cs atom contributes progressively less charge than Li due to their lower electronegativities, resulting in charge transfers of 0.71, 0.70, 0.68, and 0.67 $e^-$, respectively. The Bader charge analysis further reveals that Pd and H atoms show slight decreases in electron gain as well, which aligns with the weaker ionic character of larger alkali metals and their role as electron donors.

\begin{figure*}
\begin{center}
\includegraphics[width=2.0\columnwidth,height=8cm, draft=false]{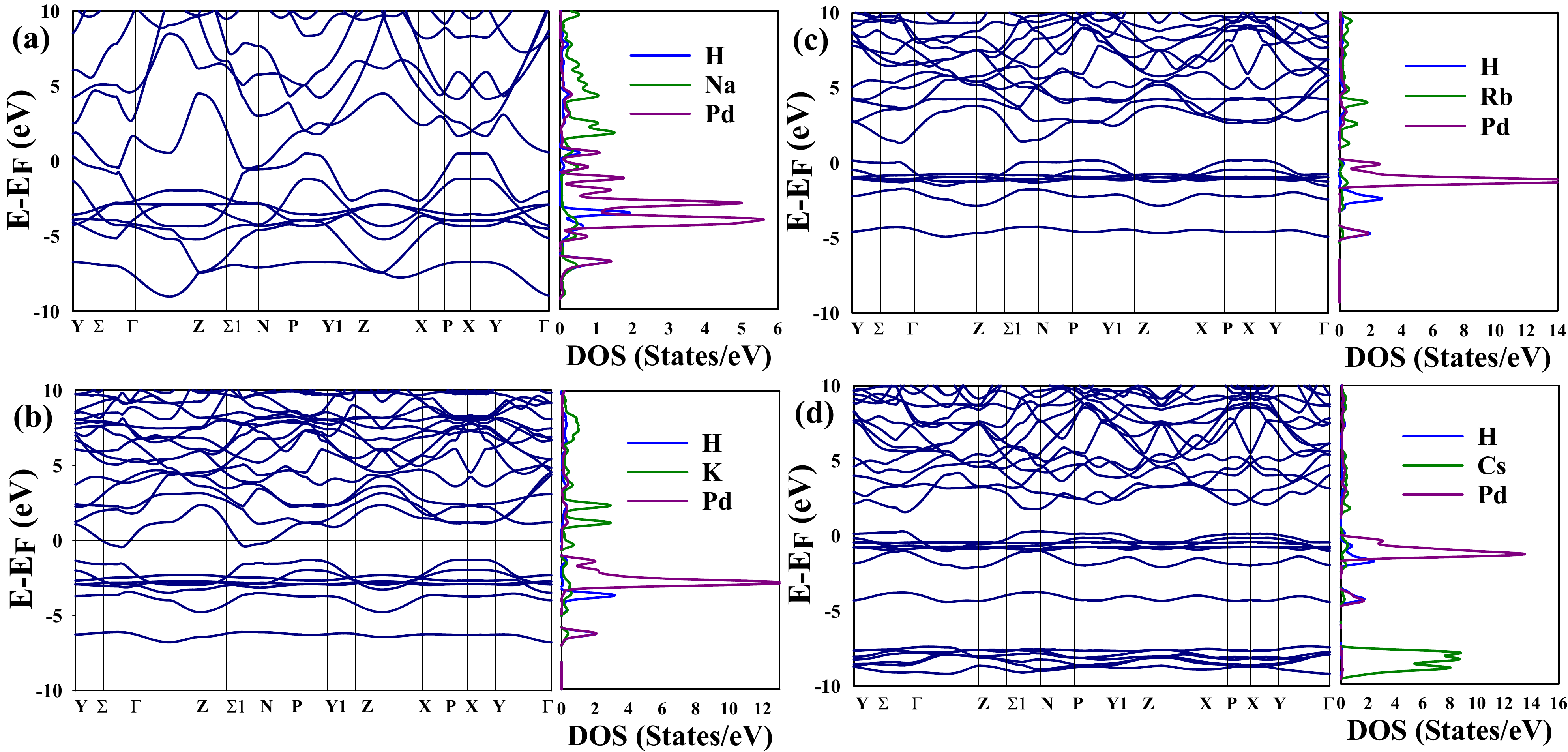}
\caption{Band structure (left panel) and density of states (DOS) (right panel) in harmonic approximation for (a) Na$_2$PdH$_2$, (b) K$_2$PdH$_2$, (c) Rb$_2$PdH$_2$ and (d) Cs$_2$PdH$_2$ structures.}
\label{Fig5}
\end{center}
\end{figure*}

The electronic band structures and DOS plots are shown in Fig. \ref{Fig5}. As the alkali metal becomes heavier, a gradual clustering of energy bands near the Fermi level is observed, particularly in Rb (Fig. \ref{Fig5}(c)) and even more prominently in Cs (Fig. \ref{Fig5}(d)). This trend indicates increasing localization of electronic states, which can be attributed to the larger ionic sizes and reduced orbital overlap in heavier alkali metal substitutions. The resulting decrease in electronic dispersion leads to narrower bands and a concentration of electronic states in specific energy regions, which may impact the metallicity and superconducting behavior.
In contrast, the bands of Na and K are almost broader, covering a wider range of energy states. This broader distribution is associated with their smaller ionic sizes and stronger orbital overlap, allowing for greater delocalization of electrons. Consequently, electrons in these bands can move more freely between atoms.

Phonon dispersion calculations were performed to assess the dynamical stability of the $A_2$PdH$_2$ compounds at ambient pressure. The results reveal that Na, K, and Rb-based structures exhibit no imaginary frequencies throughout the Brillouin zone, confirming their dynamical stability. In contrast, Cs$_2$PdH$_2$ exhibits low-frequency phonon modes with small imaginary components, which may indicate a tendency toward dynamical instability under ambient conditions. This instability is likely a result of the large ionic radius of Cs, which introduces lattice strain and weakens interatomic interactions. These findings suggest that while the $A_2$PdH$_2$ framework can support heavier alkali metals up to Rb, further stabilization—e.g., via applied pressure or structural optimization—may be required for Cs-containing variants.  However, these modes are close to zero and may be stabilized by anharmonic effects and further investigation is required to clarify the dynamical behavior.

To assess the superconducting potential of $A_2$PdH$_2$ compounds, we performed electron–phonon coupling (EPC) calculations for the dynamically stable compositions within this family. Both Na$_2$PdH$_2$ and K$_2$PdH$_2$ exhibit moderate EPC strength (λ = 0.38), but their critical temperatures differ due to disparities in phonon energy scales. Specifically, the higher logarithmic average phonon frequency ($\omega_{\log}$) in Na$_2$PdH$_2$ yields a $T_c$ of approximately 3.2 K, while the softer phonon modes in K$_2$PdH$_2$ lower its $T_c$ to around 2.1 K. Although Rb$_2$PdH$_2$ exhibits the highest electronic density of states at the Fermi level, its electron-phonon coupling constant $\lambda$ is significantly low (0.14), resulting in the suppression of superconductivity ($T_c \approx 0$~K). In contrast, Na$_2$PdH$_2$ and K$_2$PdH$_2$ exhibit comparable $\lambda$ values (0.38), but the higher average phonon frequencies in Na$_2$PdH$_2$ lead to a high superconducting transition temperature. The reduced DOS at the Fermi level in Cs, along with a wide energy region with no available states just above E$_F$, may indicate a weak metallic character. For Cs$_2$PdH$_2$, significant phonon instabilities at ambient pressure prevent a reliable $T_c$ estimate. These results emphasize the sensitivity of superconducting behavior in Pd-based ternary hydrides to subtle changes in both electronic and vibrational properties, with Na$_2$PdH$_2$ emerging as the most promising compound under ambient pressure.

\begin{figure*}
\begin{center}
\includegraphics[width=2.0\columnwidth,draft=false]{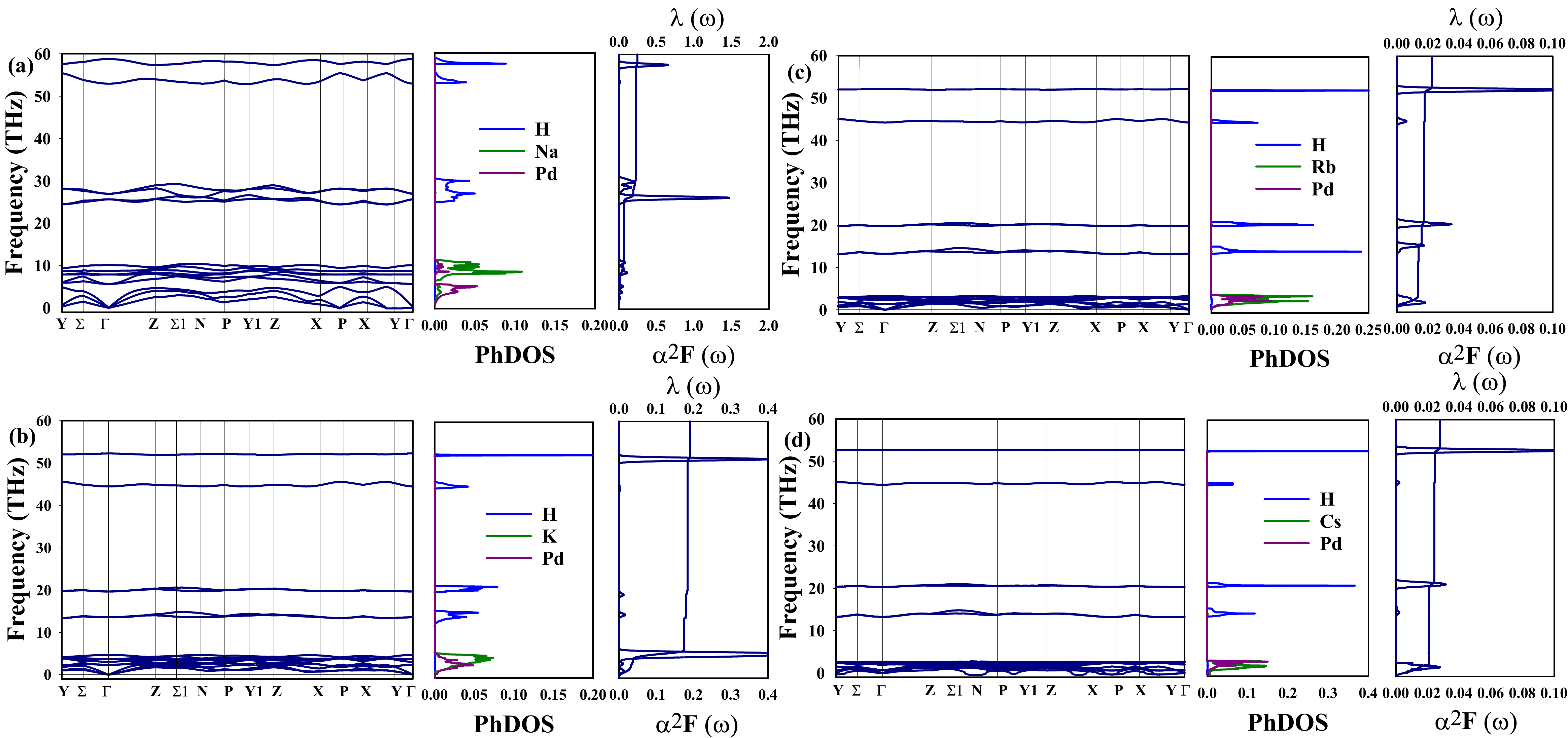}
\caption{The phonon spectrum (left), the phonon density of states (PDOS) (middle), and Eliashberg function $\alpha^{2}F(\omega)$ and integrated electron-phonon coupling constants $\lambda(\omega)$ (right) for (a) Na$_2$PdH$_2$, (b) K$_2$PdH$_2$, (c) Rb$_2$PdH$_2$ and (d) Cs$_2$PdH$_2$.}
\label{Fig6}
\end{center}
\end{figure*}

\begin{table}
\begin{center}
 \begin{tabular}{|M{1.5cm}|M{1cm}|M{1cm}|M{1cm}|M{1.5cm}|M{1.9cm}|}
 \hline
 Compound & $\lambda$ & $\omega_{log}$ & $N_F$ & $T_c$ (McMillan) & $T_c$ (Allen-Dynes) \\ \hline
 Na$_2$PdH$_2$ & 0.38 & 82.6 & 0.41 & 3.1 & 3.2 \\ 
 K$_2$PdH$_2$ & 0.38 & 54 & 0.42 & 2 & 2.1 \\    
 Rb$_2$PdH$_2$ & 0.14 & 35.6 & 1.95 & - & - \\ \hline
 \end{tabular} 
\caption{Electron-phonon coupling constant $\lambda$, logarithmic average of the phonon frequencies $\omega_{log}$ (in meV), density of states at the Fermi
level $N_F$ (in states/eV. f.u.), transition temperatures $T_c$ (in K) with the Coulomb pseudopotential parameter ${\mu}^*$ of 0.1.} 
\label{tab1}
\end{center}
\end{table}

\section{Conclusions}
In this study, we systematically investigated the structural stability, vibrational properties, and superconducting behavior of Li$_2$PdH$_2$ under varying pressures. Using a combination of structure prediction and first-principles calculations, we identified a pressure-induced phase transition from a tetragonal $I4/mmm$ phase, stable at ambient conditions, to a monoclinic $C2/m$ structure that becomes thermodynamically favorable above 5 GPa and remains stable up to at least 50 GPa. This transition is consistent with previous experimental findings and validates our computational approach. Phonon and electron–phonon coupling analyses reveal that only the high-pressure $C2/m$ phase shows superconductivity, with a $T_c$ increasing from 0.6 K at 10 GPa to 4.7 K at 50 GPa. In contrast, the low-pressure $I4/mmm$ structure remains non-superconducting, at both harmonic and anharmonic phonon corrections.

Following the identification of metallic behavior in Li$_2$PdH$_2$, we extended our analysis to the isostructural series $M_2$PdH$_2$ (M = Na, K, Rb, Cs) to assess the impact of alkali-metal substitution on structural, electronic, and superconducting properties.

Our results indicate that Na, K, and Rb-based compounds are dynamically stable at ambient pressure and preserve the [PdH$_2$] structural motif, with gradual lattice expansion in line with the increasing ionic radius. Electronically, heavier alkali substitutions lead to increased localization and reduced density of states at the Fermi level, particularly in Cs. Phonon calculations confirm dynamical stability for Na, K, and Rb, but reveal pronounced instabilities in Cs$_2$PdH$_2$ in the harmonic approximation, likely due to excessive lattice strain. Electron–phonon coupling calculations further show weak  superconducting behavior in Na$_2$PdH$_2$ ($T_c \sim$ 3.2 K) and K$_2$PdH$_2$ ($T_c \sim$ 2.1 K), with significantly weaker coupling in Rb and dynamical instability precluding $T_c$ estimation in Cs. These results highlight Na$_2$PdH$_2$ as the most promising candidate for superconductivity among the series under ambient pressure.

\section{ACKNOWLEDGMENTS}
This work is based upon research funded by Iran National Science Foundation (INSF) under project No. 4003531. Partial financial support by the Research Council of the University of Tehran is acknowledged. 
I.E. acknowledges funding from the Department of Education, Universities and Research of the Eusko Jaurlaritza, and the University of the Basque Country UPV/EHU (Grant No. IT1527-22) and 
the PID2022-142861NA-I00 project funded by MICIU/AEI/10.13039/501100011033 and FEDER, UE.
This work, in the part contributed by Y.-W. F. and I.E., was also partially supported by the IKUR Strategy-High Performance Computing and Artificial Intelligence (HPC$\&$AI) 2025-2026 of the Department of Science, Universities and Innovation of the Basque Government.

\bibliographystyle{unsrt}
\bibliography{reference}
%

\end{document}